\documentstyle[aps]{revtex}
\begin{document}
\input{psfig.sty}
\title{ On the  $\eta $ and f$_1$(1420) Couplings   to the Nucleon}
\author{S. Neumeier$^1$ and M. Kirchbach$^{2,3} $\\
$^1${\small \it  Institut f\"ur Theoretische Physik, 
Universit\"at Leipzig, 
D-04109 Leipzig, Germany}\\
$^2${\small \it Institut f\"ur Kernphysik, Universit\"at Mainz,
D-55099 Mainz, Germany\\
$^3$Escuela de Fisica de la UAZ, AP C-600, Zacatecas, ZAC 98068 Mexico
 } }
\maketitle
\begin{abstract}
We consider neutral pseudoscalar, $\eta $, and axial vector,
$f_1(1420)$, mesons in the OZI-rule-respecting flavor basis, 
$\lbrace (\bar s s),\, {1\over \sqrt{2}}(\bar u u + \bar d d)\rbrace$, 
 and suggest a scenario for their coupling to the nucleon. 
Within this framework, the non--strange parts of the   
$\eta N$ and f$_1 N$ couplings 
are modeled by means of triangular $a_0\pi N$, and $K K^* (\Lambda /\Sigma )$ 
vertices, while the strange ones partly 
proceed via Goldberger-Treiman relations, 
which have been concluded solely on the grounds of
current universality.
The suggested model explains the observed suppression 
of the $\eta N$ coupling with respect to the constituent quark 
model expectations, and predicts the coupling of $f_1$ to the nucleon.

\end{abstract}


\section{Introduction}

The present paper devotes itself to the study of
couplings of neutral 
pseudoscalar and axial vector mesons,  such as 
$\eta $ and $f_1(1420)$, 
to the nucleon in the OZI-rule-respecting
flavor basis: 
$\lbrace (\bar s s),\, {1\over \sqrt{2}}(\bar u u +\bar d d)
\rbrace$. In the process, the surprising smallness of 
the $\eta N$ coupling observed in various reactions 
finds a natural explanation. This problem has recently been 
reviewed in Ref.~\cite{KiWe}.

In the considered framework, the $(\bar s s)$ components of the
$\eta $ and $f_1$(1420) wave functions,
couple to the nucleon partly at tree-level via a Goldberger-Treiman 
(GT) relation that follows from the universality of the 
strange axial quark current, and partly via the gluon axial current.
On the contrary, within the context of current universality,
tree-level ${1\over \sqrt{2}}(\bar u u +\bar d d)N$
couplings are ruled out as the weak axial current does not contain the
light flavors in a symmetric combination \cite{KiWe}.  
We here take a pragmatic position and mimic such vertices by    
means of triangular corrections. To be specific,
we first  bridge the initial ${1\over \sqrt{2}}(\bar u u + \bar d d)$
quarkonium to the meson cloud surrounding the nucleon and then couple 
the secondary mesons to universal currents. For example, the 
${1\over \sqrt{2}}(\bar u u +\bar d d )$ part of the $\eta $  wave 
function  (to be denoted by $\eta ^q$) can be linked to 
the charged nucleon vector and axial vector currents by means of 
$a_0\pi N$ triangular vertices. Indeed, the $a_0$ meson picks up the 
isotriplet vector current, while $\pi $ selects the isotriplet axial one. 
In other words, we approximate  
$\langle \eta^q N|J_{\mu, 5}^q|N\rangle $ by 
$\langle \eta \pi|a_0\rangle \langle a_0N|N\rangle \langle N\pi | N\rangle $.
Similarly, the ${1\over \sqrt{2}}(\bar u u +\bar d d)$ part  
from the f$_1$ state (to be denoted by $f_1^q$) is coupled 
to the nucleon by means of  $K K ^*(892)(\Lambda /\Sigma ) $ 
triangles. In due course, the  couplings of 
the neutral pseudoscalar and axial vector mesons to the nucleon 
arise as combinations of the small fraction of 
nucleon helicity carried by the strange quark sea,
and the small triangular vertex corrections. It is not surprising, therefore, 
that $g_{\eta NN}$ appears suppressed relative to quark model predictions 
based on the octet Goldberger--Treiman relation.   
In this way, instead of interpreting the smallness of the
$\eta N$ coupling in terms of a suppressed 
nucleon octet axial matrix element\footnote{Note, that the problem of
the suppression of the axial octet-nucleon
coupling has been addressed in  Ref.~\cite{Sav} within
the framework of SU(6) chiral perturbation theory to one-loop order . 
There, the strong dependence of the result on the $N$--$\Delta $ mass 
splitting was revealed and the necessity for higher-order corrections 
discussed.}, we here rather reformulate this
problem in terms of an enhancement of axial $ ( \bar s s )N$ 
couplings due to vertex corrections.

A method similar in spirit to the one presented here but of different
techniques is the so--called meson cloud model of Ref.~\cite{Szczurek}.
There, the authors study the influence of the mesons surrounding the
nucleon on its electromagnetic and weak vector form factors 
in terms of a nucleon wave function having a 
non--negligible overlap with the nucleon-meson scattering continuum.
We here focus rather on the isoscalar axial form factor.

The paper is organized as follows. In the next section 
we briefly outline the calculation scheme of  
$\eta NN$ and f$_1$(1420)$NN$ vertices in the flavor-basis.
In Sect.\ 3 the effective $\eta NN$ vertex is calculated 
{}from the $\pi a_0(980)N$ triangular vertex, while Sect.\ 4 contains
the results on the effective f$_1$(1420) meson nucleon-couplings associated 
with the $KK^*(892) Y$ triangle. The paper ends with a short summary.

\section{The  $\eta $(f$_1)N$ coupling in the flavor basis }

The $\eta $ wave function in the 
$\lbrace (\bar s s),\, {1\over \sqrt{2}}(\bar u u +\bar d d)\rbrace $
flavor basis is parametrized as \cite{Feld}:
\begin{eqnarray}
|\eta \rangle =\cos \theta \,  |\eta^q\rangle &-&\sin \theta 
                \, |\eta ^s\rangle 
\, ,\nonumber\\
\cos\theta =0.773\, , &\quad & \sin\theta =0.634\, ,\nonumber\\
\eta ^q = {1\over \sqrt{2}}(\bar u u +\bar d d)\, ,
&\quad & \eta ^s =\bar s s\, .
\label{eta_OZI}
\end{eqnarray}
In the following  we assume the pseudoscalar
$\eta  (\eta ' ) N$ coupling $g_{\eta (\eta ') NN}$ to be patterned after the 
states flavor mixing
\begin{eqnarray}
g_{\eta NN} &= & g_{\eta^q NN}\, \cos\theta  -  
g_{\eta^s NN}\, \sin\theta \, ,
\label{getaN_OZI}\\
g_{\eta' NN} &=&g_{\eta^q NN}\, \sin\theta  +  
g_{\eta^s NN}\, \cos\theta \, ,
\label{getaNPr_OZI}
\end{eqnarray}
In order to obtain  $g_{\eta^s NN}$, the pseudoscalar coupling of
the $0^-$ strange quarkonium to the nucleon,
one may exploit Sakurai's idea on current universality and
field-current identity to relate the tree-level
part of $g_{\eta^s NN}$ to the weak decay constant $f_{\eta^s}$ via
a Goldberger-Treiman type relation 
\begin{eqnarray}
g_{\eta^s NN}^{GT}& = & {{g_A^s} \over f_{\eta^s}}m_N\, .
\label{etas_GT}
\end{eqnarray}
Here, $m_N$ and $g_A^s$ are in turn the nucleon mass and the
strange axial coupling.
The GT-relation in Eq.~(\ref{etas_GT}) can be interpreted
as a consequence of  ($\bar s s$)-pole dominance of the strange 
quark axial current (see Fig.\ 1). 
The net $\eta N$ coupling contains in addition a term, here denoted
by $g^R_{\eta ^s NN}$, and which is associated with 
the pseudoscalar $\eta^s (\widetilde{G}G)$ coupling,
\begin{eqnarray}
{\cal V}^R_{\eta ^s NN} &=& g_{\eta ^s NN}^R m_{\eta'}^2{f_{\pi^0} }\, 
\bar {\cal U}_N\,(\vec{p}\, ') \widetilde{G}G\,  
{\cal U}_N\,(\vec{p}\, )\phi_{\bar s s} \, .
\label{gluon_axcurr}
\end{eqnarray}
Here, ${\cal U}_N\, (\vec{p}\, ) $ is the nucleon Dirac spinor,
$G$ is the gluon field, $\widetilde{G}$ is its dual, 
$\phi_{\bar s s}$ denotes the strange quarkonium field.
The $\eta '$ mass $m_{\eta '}$ and the pion weak decay constant
$f_{\pi^0} $ serve as scale parameters.
In other words, $g_{\eta NN}$ decomposes into
\begin{eqnarray}
g_{\eta^s NN} &=&g_{\eta^s NN}^{GT} -g_{\eta^s NN}^R \, ,
\label{etas_NN} 
\end{eqnarray}
A rough estimate for $g^R_{\eta ^s NN}$ can be obtained from
Shore--Venziano's expression \cite{Ven} for $g_{\eta ' NN} $, 
the pseudoscalar $\eta ' N$ coupling, 
\begin{equation}
g_{\eta 'NN} = g_{\eta' NN}^{GT} -g^R_{\eta ' NN} \, .
\label{etapr_PS}
\end{equation}
Here, $g_{\eta ' NN}^R = {1\over \sqrt{3}}f_{\pi^0} m_{\eta '}^2 g_{GNN}$,
with $g_{GNN}$ standing for the  $(\widetilde{G} G) N$ coupling.
With the numerical values for $g_{\eta ' NN} \approx $ 1, 
and $g_{\eta ' NN}^{GT}=1.67 $ deduced from \cite{Feld1} after
accounting for the additional factor of $1/\sqrt{2}$ in their current
normalizations, we obtain $g_{\eta ' NN}^R\approx 0.67$.
In noticing that 
\begin{equation}
\sqrt{3}g^R_{\eta' NN}=g^R_{(\bar u u)NN}
+ g^R_{(\bar d d)NN} + g^R_{(\bar s s)NN}
\label{gluon_coupl}
\end{equation}
and assuming $g^R_{(\bar q_i q_i)NN}$ to be flavor independent,
we estimate $g^R_{\eta^s NN}\approx 0.4 $.

On the other hands, the strange axial coupling $g_A^s$ that 
enters the strange GT relation equals $\Delta s$, the genuine 
quark part of $a_s$, the fraction of nucleon 
polarization carried by the strange quark-gluon sea. 
In the notation of Ref.~\cite{proton}
$a_s$ is given by
\begin{equation}
a_s (Q^2) =\Delta s  -{{\alpha_s(Q^2)}\over {2\pi }}\Delta g(Q^2)\, .
\label{sSpin_frac}
\end{equation}

In using the values of $a_s$(10 GeV$^2) = - 0.10 \pm $0.02,
$\alpha_s$(10~GeV$^2$)=0.25, and ~$\Delta~g~$(10~GeV$^2$)=2$\pm~$1.3
from Ref.~\cite{proton},
one finds $ \Delta s =-0.02$. We will consider this value as
compatible with zero. In the following, $g_{\eta^s NN}^{GT}$ 
will be neglected, so that
\begin{equation}
g_{\eta^s NN} \approx- g_{\eta^s NN}^R\, \approx  - 0.4\, .
\label{etas_PSC}
\end{equation}
This is the quantity that we will keep in Eq.~(\ref{getaN_OZI})

In general, pseudoscalar meson-nucleon couplings, $g_{M NN}$, are
related to gradient couplings, $f_{M NN}$, 
via the well known equality \cite{KiTi} 
\begin{equation}
{g_{M NN}\over {2m_N }}={f_{MNN}\over m_M}\, ,
\label{equiv_rel}
\end{equation}
with $m_M$ standing for the meson mass. 
That the GT-relation from Eq.~(\ref{etas_GT}) entirely originates 
from (strange) axial current universality is expressed by means of  
the following current-current couplings: 
\begin{eqnarray}
{\cal V}^s_{\eta^s N N } =  {1\over {f^2_{\eta^s}} } \, 
                       J^{s\, (N)}\cdot J^{ (\eta^s )}\,  
&=& {1\over {f_{\eta^s} ^2 }} g_A^s(Q^2)\, \bar{\cal U}_N(p')\gamma\gamma_5 
{{1\!\!1}\over 2} {\cal U}_N(p) \,\cdot \, f_{\eta^s} \, iq \phi_{\eta^s}\, ,
\nonumber\\
{\cal V}^s_{f^s_1 N N } =  {1\over m_{f^s_1}^2 } \, 
                       J^{s\, (N)}\cdot J^{ (f_{f^s_1} )}\, 
&=& {1\over m_{f^s_1}^2 } g_A^s(Q^2)\, \bar{\cal U}_N(p')\gamma\gamma_5 
{{1\!\!1}\over 2} {\cal U}_N(p) \,\cdot \, f_{f_1^s} \,m_{f^s_1}^2\, 
\phi_{f^s_1}\, .
\label{vert_etaNN}
\end{eqnarray}
Here, $\phi_{\eta^s} $ and $\phi_{f^s_1}$ are the respective
$\bar s s$ parts of the f$_1$ and $\eta $ fields, while
$J^{\eta^s} $ and $J^{f^s_1}$ are the associated axial currents.
The combination $g_A^sf_{f^s_1} /2$ is identified with 
the f$^s_1$(1420)$N$ contact coupling $f_{f^s_1NN}$,
while  $g_A^s /2f_{\eta^s} $ is associated with the gradient
${\eta^s} N$ coupling, i.e. one defines
\begin{eqnarray}
{f_{\eta^s NN} \over m_{\eta^s} } ={ {g_A^s}\over {2f_{\eta^s} } }\, , \quad
&\quad & 
f_{f_1^s NN}  ={ {g_A^s f_{f^s_1} }\over 2 }\, , \quad \mbox{with}\quad
g_A^s = \Delta s\, .
\label{etaN_vert}
\end{eqnarray}
Indeed, Eq.~(\ref{etas_GT}) is reproduced in 
inserting the first of Eqs.~(\ref{etaN_vert}) into (\ref{equiv_rel}).

As long as at tree level the  contact $\eta N$ coupling 
appears proportional to $\Delta s$ rather than follows the octet 
GT-relation $f_{\eta_8 NN}/ m_{\eta_8} = g_A^8/ (2f_{\eta_8} ) $ of 
Ref.~\cite{deAlf}, the $\eta $ meson can be interpreted as a ``partial'' 
strange Goldstone boson, a result already conjectured in a 
previous work \cite{KiWe}. Note, that the accuracy of GT--relations has 
been reliably proven only for the case of the pion--nucleon system 
\cite{Coon}.   
The subtle reason for this situation can be understood if we note that
mesons transporting weak interaction are not same as such 
participating strong interaction. This is due to the fact that the 
electroweak gauge group SU(2)$_L\otimes $U(1) and the strong flavor 
group SU(3)$_F$ do not share, in general, common multiplets. Because of 
that, the weak interaction is incapable to recognize the $\eta $ meson 
as a particle different but a strange quarkonium. 
The mismatch between the flavor contents of strong and
electroweak spin-0 mesons has been studied extensively, for example, 
in Ref.~\cite{BMa} by means of F-spin breaking through the electroweak 
gauge symmetry. Using the genuine electroweak meson states from
\cite{BMa} in the construction of weak nucleon-nucleon potentials is an
interesting goal for future research as it might 
modify interpretation of parity violation  effects in nuclei.
Now, due to the smallness of $\Delta s$, 
the $\eta$ and f$_1$(1420) mesons practically decouple at tree level 
from the nucleon.
This might be one of the main reasons for which a strong suppression of the 
$\eta N$ couplings has frequently been found over the years by various 
data analyses of $\eta $ photo-production off proton at threshold 
\cite{TiKa}, as well as nucleon-nucleon (NN) and nucleon--hyperon (NY) 
phase shifts \cite{Reuber}.
Same observation applies to $f_{f_1^s}$. In the following, 
the strange tree level couplings will be of minor importance
and  neglected.

Compared to $g_{\eta sNN}$, the calculation of the pseudoscalar 
$\eta^q NN$ coupling can not be discussed in terms of universality 
of the non-strange isosinglet axial current, as the weak axial current 
does not contain the light flavors in a symmetric combination.
Therefore,  no obvious GT-relation can be written for 
$g_{\eta^q NN}$ (see Fig.\ 2).

In the following two sections we consider the 
$\eta ^q NN$ and $f_1^q NN$  vertices to proceed entirely
via small but non--negligible triangular loops of the type 
$a_0 \pi N$ in Fig.\ 4 , and $KK^* (\Lambda /\Sigma )$ in Fig.\ 7,
 respectively.

\section{Effective $\eta NN$ Vertices}
\label{sec-etaNN}
In this section we calculate the gradient and pseudoscalar 
coupling constants of the $\eta $ meson to the nucleon by means of triangular
vertices involving the $a_0$(980) and $\pi $ mesons.

The special role of the $a_0(980)\pi N$ triangular diagram
as the dominant one--loop mechanism for the $\eta N$ coupling 
is emphasized by the circumstance that the $a_0(980)$ meson is
the lightest meson with a two particle decay channel
containing the $\eta $ particle \cite{PDG94}.

The contributions of heavier mesons such as
the isotriplet $a_2(1320)$ tensor meson with an $\eta\pi $ 
decay channel and the
isoscalar $f_0(1400)$, $f^\prime _2(1525)$ and $f_2(1720)$ tensor
mesons with $\eta \eta $ decay channels
will be left out of consideration
because of the short range character
of the corresponding triangle diagrams on the one side,%
\footnote{The same argument applies to
the neglect of the $f_0(1590)\eta N$ triangular
vertex.}
and because of the comparatively small couplings of the tensor
mesons to the nucleon \cite{Ma89,Els87} on the other side.
The $\pi a_0 N $ triangular couplings in (Fig.\ 3)
have been calculated using the following effective Lagrangians:
\begin{eqnarray}
{\cal L}_{a_0\eta \pi }(x) & = & f_{a_0\eta \pi } 
{{m_{a_0}^2 -m_\eta^2}\over m_\pi}\phi_\eta^\dagger (x) 
\vec{\phi_\pi}(x)\cdot
\vec{\phi}_{a_0}(x) \\
{\cal L}_{\pi NN}(x) & = & 
\frac{f_{\pi NN}}{m_\pi}\bar N (x)\gamma_\mu \gamma_5
\vec \tau N (x)\cdot \partial^\mu\vec {\phi}_\pi (x) , \\ 
{\cal L}_{a_0NN}(x)& = & g_{a_0 NN}\,i \bar N (x)
\vec\tau N (x)\cdot\vec{\phi}_{a_0}(x) \, .
\label{a0pi_larg}
\end{eqnarray}

Here $f_{\pi NN}$ and $g_{a_0NN}$ 
in turn denote the pseudovector $\pi N$ 
and  the scalar  $a_0N$ coupling constants. We
adopt for $f_{\pi NN}$ the standard value $f_{\pi NN}^2/4\pi = 0.075$ 
and fit  $g_{a_0 NN}$ to data. 
The value of $f_{a_0\eta \pi }=0.44$ has been extracted 
{}from the experimental decay width \cite{PDG94} when ascribing the total 
$a_0$ width to the $a_0\to \eta + \pi$ decay channel.
Note that ${\cal L}_{a_0\eta \pi }$ represents only a part of the full
chiral Lagrangians of the $a_0\eta \pi $ system constructed
in Ref.~\cite{Ecker}. The term containing the gradient $\eta \pi $ coupling
has not been taken into account so far as we restrict ourself to
Lagrangians containing a minimal number of derivatives to ensure convergence
of the expressions.  Nonetheless,
through the use of a gradient $\pi N$ coupling, the triangular 
$\eta NN$ vertex vanishes in the chiral limit despite the fact that
the $a_0\pi \eta $ vertex does not.
In the following we will identify  the nucleon Born term 
$\langle \pi N|N\rangle \langle N|a_0 N\rangle $ entering 
the diagram in Fig.\ 4 with the full amplitude 
$T_{\eta N}(\pi +N\to a_0+N )$, and  parameterize it in terms 
of the following complete set of invariants
\begin{equation} 
T_{\eta N}(\pi +N\to a_0 +N ) 
= \bar {\cal U}_N(\vec{p}\, ') \left( {{G_1(k^2)}\over m_\eta }\, 
k\!\!\! / \gamma_5 + G_2(k^2)\, \gamma_5 \right)\, {\cal U}_N(\vec{p}\, )\, , 
\label{etaN_ampl}
\end{equation}
where the invariant functions $G_1 (k^2)$ and $G_2(k^2)$ in turn correspond 
to pseudovector (PV) and pseudoscalar (PS) types of the $ \eta N$ coupling.  
Expressions for $G_1(k^2)$ and $G_2(k^2)$ can be found in evaluating the 
diagrams in Fig.\ 4 in accordance with the standard Feynman rules.
We here systematically consider the incoming proton to be on its mass
shell and make use of the Dirac equation, so that
\begin{equation}
\gamma_5 p\!\!\! / \,\, {\cal U}_N (\vec{p}\, )= m_N
\gamma_5\, {\cal U}_N(\vec{p}\, )\, ,
\label{pin_on2}
\end{equation}
holds. On the contrary, the outgoing proton has been considered 
to be off its mass shell with
\begin{equation}
\gamma_5 p\!\!\!/\, '  {\cal U}_N(\vec{p}\, ) =
\gamma_5 (p\!\!\!/+k\!\!\!/){\cal U}_N(\vec{p}\, )
=m_N\gamma_5{\cal U}_N(\vec{p})+
 \gamma_5k\!\!\!/\, {\cal U}_N(\vec{p}\, )\, .
\label{pout_off2}
\end{equation}
The final result on $G_1(k^2)$ obtained in this way reads:
\begin{eqnarray}
G_1(k^2) & = &
C\, \int_0^1\int_0^1 dy dx x {{c_1 (x,y,k^2)}\over
  { {\cal Z} (m_N, m_\pi,m_{a_0}, x,y,k^2 )} }\, , \nonumber\\
c_1(x,y,k^2)& =  & -{1\over 2} x(1-y)m_Nm_\eta  \, , \nonumber\\
C &=& \frac{3}{8\pi^2}{{m_{a_0}^2-m_\eta^2}\over m_\pi^2}
 f_{\pi NN}f_{a_0\eta \pi}g_{a_0 NN} \, . 
\label{f_eta}
\end{eqnarray}
The corresponding expression for $G_2(k^2)$ reads
\begin{eqnarray}
G_2(k^2) & = &
C\, \int_0^1\int_0^1 dy dx x {{c_2(x,y, k^2)}\over
{{\cal Z}(m_N, m_\pi,m_{a_0}, x,y,k^2)}}\, ,\nonumber\\
c_2(x, y, k^2) & =& -x(1-y)m_N^2 \, .
\label{g_PS} 
\end{eqnarray}
The function ${\cal Z}(m_B,  m_1,m_2, x,y,k^2)$  appearing in the last
two expressions is defined as
\begin{eqnarray}
{\cal Z}(m_B, m_1,m_2, x,y, k^2) & =& 
m_N^2x^2 (1-y)^2 +  x^2yk^2 + m_1^2(1-x) + (m_2^2- k^2)xy\nonumber\\ 
&+&(m_B^2-m_N^2)x(1-y)\, .
\end{eqnarray}

The remarkable feature of the analytical expressions for the
pseudoscalar and pseudovector $\eta N$ couplings is that
they are given by {\it completely convergent\/} integrals and depend
only on the $a_0\to \pi +\eta $ decay constant and the
respective pion and $a_0$ meson-nucleon couplings.
The sources of uncertainty in the parameterization of the
effective $\eta NN $ vertex by means of
the triangular $a_0(980)\pi N$ diagram are
associated with the $a_0(980) N$ coupling constant and
the $\Gamma (\eta \pi)/ \Gamma^{\rm tot}_{a_0}$
branching ratio. For example, the coupling constant $g_{a_0NN}$ varies 
between $\approx 3.11$ and $\approx 10$ depending on the $NN$ potential 
model version \cite{Ma89,Els87}. Because of that we give below
the values for the gradient and pseudoscalar $\eta N$ couplings 
following from the $a_0\pi N$ triangular $\eta N$ vertex 
as a function of the $a_0N$ coupling constant:
\begin{eqnarray}
|G_1 (k^2 =m_\eta^2)|  = 0.06 \, g_{a_0NN}\, , &\quad & 
|G_2 (k^2 =m_\eta^2 )| = 0.22 \, g_{a_0NN}\, .
\label{eff_a0pi}
\end{eqnarray}
There are the quantities in Eqs.~(\ref{f_eta}) and (\ref{g_PS})
which we shall interpret as the {\it effective\/}
 pseu\-dovec\-tor and pseudoscalar
$\eta^q N$ coupling constants, respectively,
\begin{eqnarray}
\cos \theta f^{\rm eff}_{\eta^q NN} (k^2) \, =\,  G_1(k^2)\, ,&\qquad &
\cos\theta \, g^{\rm eff} _{\eta^q NN} (k^2) \, = \, G_2 (k^2)\, ,
\quad k^2 =m_\eta^2 .
\label{PVPS_cpl}
\end{eqnarray}
In combining the last equation with (\ref{getaN_OZI}), 
and (\ref{etas_PSC}), one finds the following expression for 
the $\eta N$ coupling:
\begin{eqnarray}
g_{\eta NN} &=& 0.22 g_{a_0 NN} + 0.63 g_{\eta^s NN}^R 
\approx 0.22 g_{a_0 NN} +0.25\, .
\label{geta_NN_expr}
\end{eqnarray}
In using four different values for $g^2_{a_0NN}/4\pi $ 
reported in \cite{Ma89}, \cite{Els87},
we calculate  $g^2_{\eta NN}/4\pi  $ in Table 1.
Our analyses suggest for the pseudoscalar $\eta N$ coupling small 
but non-vanishing values. In comparing our predictions to data, we
rely upon Refs.~\cite{TiKa}. There, Tiator {\it et al.} interpreted the 
high-precision MAMI measurement of the differential cross sections in 
$\eta $ photo-production off--proton near threshold in terms of the strongly 
suppressed $g_{\eta NN}$ value of $g_{\eta NN}^2 /4\pi \approx 0.4$. 
This result was concluded from the small $P$ wave contribution 
to the almost flat angular distributions for a wide range of beam energies.
Note, that a larger value for $g_{\eta NN}$ has been extracted 
from total $\bar p p $ cross sections in using dispersion
relation techniques \cite{GrKr}. 
However, the latter procedure is endowed with more 
ambiguities than the straightforward calculation of the angular 
distributions from Feynman diagrams presented in Ref.~\cite{TiKa}. 

{}From Eq.~(\ref{etaN_ampl}) one sees that the triangular $a_0\pi N$ 
correction to the $\eta NN $ vertex represents a mixture
\cite{Gross} of pseudovector (PV) and pseudoscalar (PS) types of $\eta N$ 
couplings. This mixing is quite important for reproducing the form of the
differential cross section for $\eta $ photo-production off proton
at threshold in Fig.\ 6.
Note that such a mixing cannot take place for Goldstone bosons 
because their point-like gradient couplings to quarks are determined
in a unique way. On the contrary, in case of extended effective 
meson-nucleon vertices, such a mixing can take place by means of 
Eq.~(\ref{etaN_ampl}). However, the PV-PS separation is model dependent
and can not be interpreted as observable.
Data compatibility with the PV-PS mixing created by the $\pi a_0 N$ triangular 
correction to the $\eta NN$ vertex is a further hint on the non--octet
Goldstone boson nature of the $\eta $ meson
\footnote{It should be noted that the analytical expressions for 
the $\eta N$ couplings 
in Eqs.~(\ref{g_PS}) and (\ref{f_eta}) differ from those 
obtained in \cite{KiTi} where the ambiguity in treating the
off--shellness of the outgoing proton was not kept minimal.
\begin{table}[htbp]
\caption{The dependence of the $\eta N$ coupling on $g_{a_0NN}$ following
from Eq.~(24)}
~\\
\begin{tabular}{lcccc}
~\\
 ${g^2_{a_0 NN}\over {4\pi }}$& 0.77 & 1.075& 3.71 & 6.79 \\ 
~\\
\hline
~\\
${g^2_{\eta NN}\over {4\pi }}$& 0.07 &0.09& 0.24 &0.41  \\
~\\
\end{tabular}
\end{table}
In the present calculation, the on--shell approximation 
$p\cdot k =-p'\cdot k = -m^2_\eta /2 $ was made
only in evaluating the denominators of the Feynman diagrams, whereas
in the nominators $p\!\!\!/ ' $ was consequently replaced by
$p\!\!\!/ ' =\, p\!\!\!/ + \, k\!\!\!/ $. In contrast to this, in \cite{KiTi}
the above on--shell approximation was applied to the nominators
too and, in addition, terms containing $p\!\!\!/ '$ have been occasionally 
interpreted as independent couplings. Through the improper treatment of the
off-shellness of the outgoing proton in \cite{KiTi}, 
logarithmically divergent integrals have been artificially invoked 
in $g_{\eta NN}$.}.
{}Finally, in estimating the $\eta' $ nucleon coupling in terms
of Eqs.~(\ref{getaNPr_OZI}), (\ref{etas_PSC}), (\ref{eff_a0pi}),
and (\ref{PVPS_cpl}) we find
\begin{equation}
g_{\eta 'NN}={\sin\theta \over \cos\theta }\, 0.22 g_{a_0 NN}
+\cos\theta \, g^R_{\eta^q NN}\, .
\end{equation}
{}For the $\eta$--$\eta ' $ mixing angle of 
$\theta \approx 39.4^\circ $
and the maximal value of $g_{a_0 NN} \approx \, 9.24 $, one finds
$g_{\eta' NN} \approx 1.41 $ and in agreement with the estimate
of Ref.~\cite{Feld1}. Thus in simulating the anomalous 
${1\over \sqrt{2}}(\bar u u + \bar d d)N$ coupling by the
triangular $a_0\pi N$ vertex we are able to explain the smallness
of both the $\eta ' N$ and $\eta N$ couplings. 
 
\section{Effective f$_1$(1420)$NN$ Vertices}
\label{sec-f1NN}
The internal structure of the axial vector meson 
f$_1$(1420) is still subject to some debates (see Note on f$_1$(1420)
in \cite{PDG94}). Within the constituent quark model this meson
is considered as the candidate for the axial meson ($\bar s s$)
state and therefore as the parity partner to $\phi $ 
{}from the vector meson nonet. 
 The basic difference between the neutral $1^-$ and $1^+$
vector mesons is that while the physical $\omega $ and $\phi $
mesons are almost perfect non-strange and strange quarkonia, respectively,
their corresponding parity partners f$_1$(1285) and f$_1$(1420)
are not. For these axial vector mesons strange and non--strange 
quarkonia appear mixed up by the angle $\epsilon \approx  15^\circ$
(compare Ref.~\cite{Bolton}):
\begin{eqnarray}
|f_1(1285) \rangle &=& -\sin\epsilon\, (|\bar s s\rangle)
+\cos\,  \epsilon\, {1\over \sqrt{2}}(|\bar u u +\bar d d\rangle)\, ,
\label{D_Weyl} \\
|f_1(1420) \rangle &=& \cos\, \epsilon\, (|\bar s s\rangle)
+ \sin\,  \epsilon\, {1\over \sqrt{2}}(|\bar u u +\bar d d\rangle)\, ,
\quad
\epsilon \approx  15^\circ\,  .
\label{E_Weyl} 
\end{eqnarray}
Therefore within this scheme the violation of the OZI rule
for the neutral axial vector mesons appears quite different as
compared to the pseudoscalar mesons.  If one had to re-write 
the $\eta $ wave function to the form of Eq.~(\ref{E_Weyl}),
one had to use $\epsilon \approx ({\pi\over 2} + 39.3^\circ ) $.
In general, as compared to pseudoscalar mesons, 
the axial vector meson sector respects better the OZI rule.
This situation is often discussed
in terms of significant glueball presence in the vector meson wave 
functions and as the consequence of the U(1)$_A$ anomaly \cite{Zou}. 
On the other side, the f$_1$(1420) meson seems 
alternatively to be equally well interpreted as a $K^*\bar K$ molecule 
\cite{Kmol}. The coupling of this meson to the nucleon is not 
experimentally well established so far. The only information about it
 can be obtained from fitting $NN$ phase shifts  by means of generalized 
boson exchange potentials of the type considered in Refs.\cite{Ma89,Els87}. 
There, one finds that the coupling $|g_{f_1 NN}| \, \approx 10$ to the 
nucleon of an effective f$_1$ meson having same mass as the 
{}f$_1$(1285) meson is comparable to the $\omega N$ coupling. 
Below we demonstrate that couplings of that magnitude can be associated to a 
large amount with triangular diagrams of the type $KK^*Y$ with 
$Y=\Lambda ,\Sigma $.

To calculate the $KK^*Y$ triangular coupling in (Fig.\ 7)
we use the following effective Lagrangians:
\begin{eqnarray}
{\cal L}_{f_1KK^* }(x) & = & f_{f_1KK^* } 
{{m_{K^*}^2 -m_K^2}\over m_K} \, 
K^\dagger (x) K^* (x)^\mu f_1 (x)_\mu + h.c.\, , \label{f1KKvertex}\\
{\cal L}_{K Y N}(x) & = & 
g_{KNY} \, i\, \bar Y (x)\gamma_5 N (x) 
 K (x) +h.c. , \\ 
{\cal L}_{K^* YN}(x)& = & 
-g_{K^*YN}\bar N (x) (\gamma^\mu K^* (x)_\mu 
+{{\kappa_V}\over {m_N+m_Y}}
\sigma^{\mu\nu} \partial_\nu K^*(x)_\mu )Y (x)   +h.c.\,  
\label{KK*_lagr}
\end{eqnarray}
Here, $g_{KNY}$ stands for the pseudoscalar coupling of the
kaon to the nucleon-hyperon ($Y$) system, $g_{K^*NY}$ denotes
the vectorial $K^* NY$ coupling, $ g_{K^*NY}{{\kappa _V}\over {m_N+m_Y}} $ 
is the tensor coupling of the $K^*$ meson to the baryon, $K^*(x)_\mu $, 
and f$_1(x)_\mu $ in turn stand for the polarization vectors of the $K^*$
and f$_1$ mesons, $K(x)$ is the kaon field, while $N(x)$ and $Y(x)$ are in turn
the nucleon and hyperon fields.

A comment on the construction of the effective 
Lagrangian ${\cal L}_{f_1KK^*}$ is in place. 
Ogievetsky and Zupnik (OZ) \cite{OZ} constructed
a chirally invariant Lagrangian containing no more than two
{}field derivatives to describe the dynamics of the 
$a_1(1260) \rho(770) \pi$ system. Exploiting the fact that these mesons
have the same external quantum numbers $J^{PC}$ as the 
respective f$_1$(1420), $K^{\ast},$ and $K$ mesons,
the OZ Lagrangian might serve as a model for
the f$_1  K^* K$ system. However, the calculation shows up
UV-divergent integrals which are to be maintained
by additional cut--offs, i.e.\  by introducing form factors 
at both the $KYN$- and $K^*YN$-vertices \cite{SN}. 
This inconvenience makes the Lagrangian choice according to 
\cite{OZ} less attractive than the present one given above
by Eq.~(\ref{f1KKvertex}). 

We adopt for the coupling constants the values implied by the J\"ulich 
potential \cite{Reuber}
\begin{eqnarray}
{{g^2_{p\Lambda K^+}}\over {4\pi}}
= (-0.952)^2\, , &\quad & {{g^2_{p\Lambda K^*}}\over {4\pi}}
= (-1.588)^2\, ,\quad \kappa_V = 4.5\, .
\label{KK*_couplings}
\end{eqnarray}
The value for $f_{f_1KK^*}= 1.97$ has been extracted from the experimental
$\Gamma_{\bar K K^*+h.c.}$ width of 17 MeV \cite{PDG94} and are compatible
with the size of the corresponding GT relations for kaons.
In identifying now the hyperon Born term
 $\langle KY|Y\rangle \langle Y|K^*N\rangle $ 
entering the effective f$_1NN$ vertex 
with the full amplitude $T_{f_1N}(K+N\to K^*+N) $, we
 expand the latter into the complete
set of the following invariants 
\begin{eqnarray}
T_{f_1N} (K+N\to K^*+N) & = & \bar {\cal U}_N(\vec{p}\, ')\,  
\Big( F_1 (k^2)\, \gamma\cdot \epsilon_{f_1}\gamma_5 +
{{F_2 (k^2)}\over m_{f_1}} \, 
\gamma \cdot \epsilon_{f_1}\gamma_5 \gamma\cdot k\, 
\nonumber\\
&+& {{F_3(k^2)}\over {m_Nm_{f_1}}}\,  
p\cdot \epsilon_{f_1}\gamma_5\gamma\cdot k 
+ {{F_4(k^2)}\over m_N}\, p\cdot \epsilon_{f_1} \gamma_5\, \Big) 
\, {\cal U}_N(\vec{p}\, )\, .
\label{f1_coupl}
\end{eqnarray}
It will become clear in due course that 
the vector part of the $K^*NY $ coupling contributes only to the 
$F_2(k^2)$ and $F_4(k^2)$ currents in Eq.~(\ref{f1_coupl}). All the remaining
terms are entirely due to the $K^*NY$ tensor coupling.
{}For small external momenta, i.e. when 
$p\!\!\!/'{\cal U}_N (\vec{p}\, ) \approx m_N{\cal U}_N(\vec{p}\, )$,   
the number of the invariants reduces to three as the 
{}first term in Eq.~(\ref{f1_coupl}) can be approximated by the 
{}following linear combinations of the $F_2(k^2)$ and $F_3(k^2)$ currents:
\begin{eqnarray}
m_N \bar {\cal U}_N (\vec{p}\,')\gamma_5 k\!\!\!/\gamma_\mu 
{\cal U}_N (\vec{p}\, )
= - p'\cdot k\, \bar {\cal U}_N(\vec{p}\, ) \gamma_5\gamma_\mu 
{\cal U}_N(\vec{p}\, )
&+& p_\mu ' \, \bar {\cal U}_N(\vec{p}\, )\gamma_5\, k\!\!\!/
\, {\cal U}_N (\vec{p}\, )\, .
\label{trick}
\end{eqnarray}
One can make use of the approximation of Eq.~(\ref{trick}) 
to estimate $F_1(k^2)$, the only invariant function which cannot be 
deduced in an unique way from the triangular $KK^*Y$ diagrams 
because of the (logarithmically) divergent parts contained there.

The invariant functions $F_i(k^2)$ are now evaluated by means of 
the diagrams of momentum--flow in 
Fig.\ 5
in using the techniques of the previous section.
The various f$_1 N$ couplings can then be expressed in terms
of the integrals
\begin{equation}
 F_i(k^2) = \frac{1}{16\pi^2} f_{f_1KK^*} g_{KYN}g_{K^*YN}\, \int_0^1\, 
x\mbox{d}x\int_0^1\mbox{d}y 
B_i(k^2)\, .
\label{f1_FF}
\end{equation}
The only divergent analytical expression is the one for
$B_1(k^2)$ 
\begin{eqnarray}
B_1(k^2) &=& 2\bar\kappa {{m_{f_1}^2-m_{K^*}^2}\over m_K}
m_N (xm_Y - 2m_N){1\over Z}\nonumber\\
&+& \bar \kappa k^2  {{m_{f_1}^2-m_{K^*}^2}\over m_K}
(xy+(1-x)(8xy+x-6)){1\over Z}
\, ,\nonumber\\
&-&2\bar\kappa {{m_{f_1}^2-m_{K^*}^2}\over m_K}
\ln { {{\cal Z}(m_Y,m_K,\Lambda_{K^*},x,y,k^2)
      {\cal Z}(m_Y,\Lambda_K,m_{K^*},x,y,k^2)}\over
     {{\cal Z}(m_Y,m_K,m_{K^*},x,y,k^2)
      {\cal Z}(m_Y,\Lambda_K,\Lambda_{K^*},x,y,k^2)} }\, .
\label{diverg_ff}
\end{eqnarray}
All the remaining invariants are convergent and given below as
\begin{eqnarray}
B_2(k^2) &=& 
{{m_{f_1}^2-m_{K^*}^2}\over m_K}
m_{f_1}[\bar\kappa (m_Y(-x^2y(1-y)+3xy-3x+2) + xym_N)+(1-x(1+y))]
{1\over Z}
\, , \nonumber\\
B_3(k^2)&=& -\bar \kappa m_{f_1}m_N{{m_{f_1}^2-m_{K^*}^2}\over m_K}
((1-y)(x^2(1+3y)-3x)+2)\, {1\over Z}
\, ,\nonumber\\
B_4(k^2) &=& -2m_N{{m_{f_1}^2-m_{K^*}^2}\over m_K}
[(x(1-y)-1)(1+\bar\kappa m_N x(1-y)) + \bar\kappa m_Y x(1+y) ] \,
{1\over Z}\, , \nonumber\\
Z&=& {\cal Z}(m_Y,m_K,m_{K^*},x,y,k^2) \, .
\label{B_is}
\end{eqnarray}
The numerical evaluation of the expressions in the last two equations
leads to the following results:
\begin{eqnarray}
F_1 (k^2=m_{f_1}^2) = -8.56 \, , &\quad &
F_ 2 (k^2=m_{f_1}^2) = 6.83\, ,\nonumber\\
F_3(k^2=m_{f_1}^2) = -3.77\, , &\quad &
F_4(k^2=m_{f_1}^2) = -2.03\, ,\nonumber\\
m_{f_1} &=& 1385.7\, {\rm MeV}. 
\label{Werte_Bis}
\end{eqnarray}
In the spirit of Eqs.~(\ref{getaN_OZI}) and~(\ref{E_Weyl}),
we here identify
$F_1(m_{f_1}^2)$ with $f_{f_1^q NN}\, \sin \epsilon $. 
These results show that all the couplings in Eq.~(\ref{f1_coupl})
are noticeable and have to be taken into account in calculating
$\eta $ and f$_1$ meson production cross sections.
Note, however that recent analyses of data on electromagnetic
strangeness production  performed in~\cite{Saghai}
revealed a surprising suppression of the $K^*YN$ vertices.
Using the vertex constants from \cite{Saghai} would reduce
the size of the f$_1 N$ couplings by at least an order of magnitude.

\section{Summary}

We developed a scenario for $\eta NN$ and f$_1 NN$ vertices, where
the  non-strange ${1\over \sqrt{2}}(\bar u u+\bar d d)$ quarkonia from 
the mesons wave functions were coupled to the nucleon by means of 
triangular vertex corrections, while the 
$(\bar s s )N $ couplings were of Shore--Veneziano's type. 
We used phenomenological Lagrangians containing minimal number
of derivatives to construct effective 
$\eta N$ and f$_1 N$ coupling strengths beyond the tree level
in terms of triangular $a_0 \pi N$, and $KK^* (\Lambda /\Sigma )$ 
diagrams, respectively.
We found all kinds of effective couplings (up to one) to be 
determined by divergenceless expressions. 
Despite that the loop corrections dominate the tree-level ones,
the net effect is a significant suppression of the couplings considered 
relative to quark model predictions. 
Remarkably, the $\eta N$ and $\eta ' N$ couplings  were found to be of
comparable size, opposite to quark model predictions and in line with
data. To be specific, the triangular $a_0\pi N$ vertex
revealed itself as a successful effective mechanism for the 
$\eta ^q N$ coupling and led to a satisfactory explanation
of the smallness of both $g_{\eta NN}$ and $g_{\eta ' NN}$.

\vspace{0.1cm}
\noindent
{\bf Acknowledgements}

We wish to thank Hans J.\   Weber for continuous supportive discussions
on the nature of the $\eta N$ coupling mechanism, Martin Reuter for
his interest and helpful remarks on particular group theoretical aspects,
and Andreas Wirzba for his comments on the 
U(1)$_A$ anomaly problems.

This work was partly supported by CONACYT, Mexico.

\begin{figure}[htbp]
\centerline{\psfig{figure=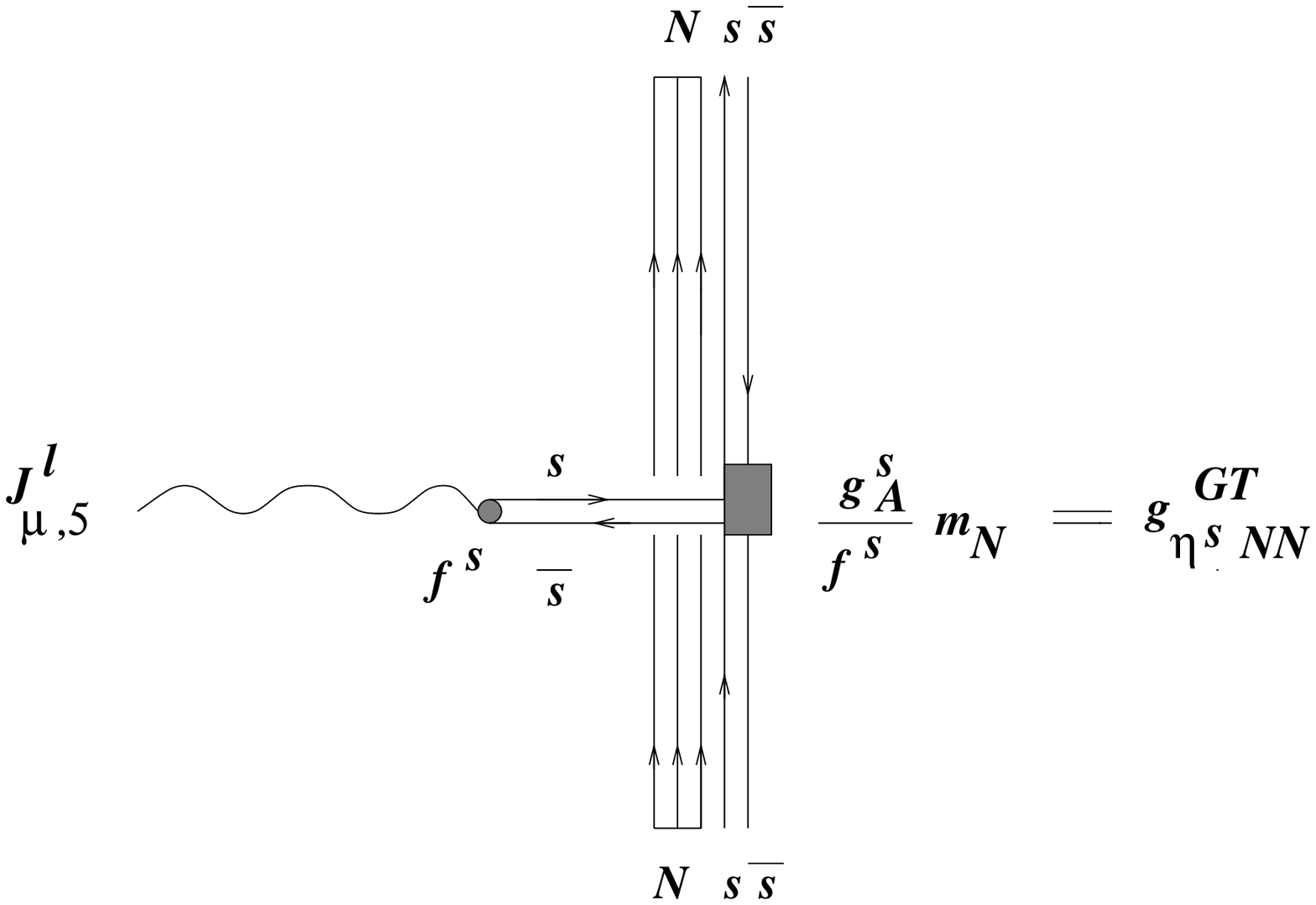,width=10cm}}
\vspace{0.1cm}
{\large Fig.\ 1\hspace{0.2cm} Universality of the strange
axial quark current and Goldberger-Treiman  $\bar s s N$ coupling. }
\end{figure} 

\begin{figure}[htbp]
\centerline{\psfig{figure=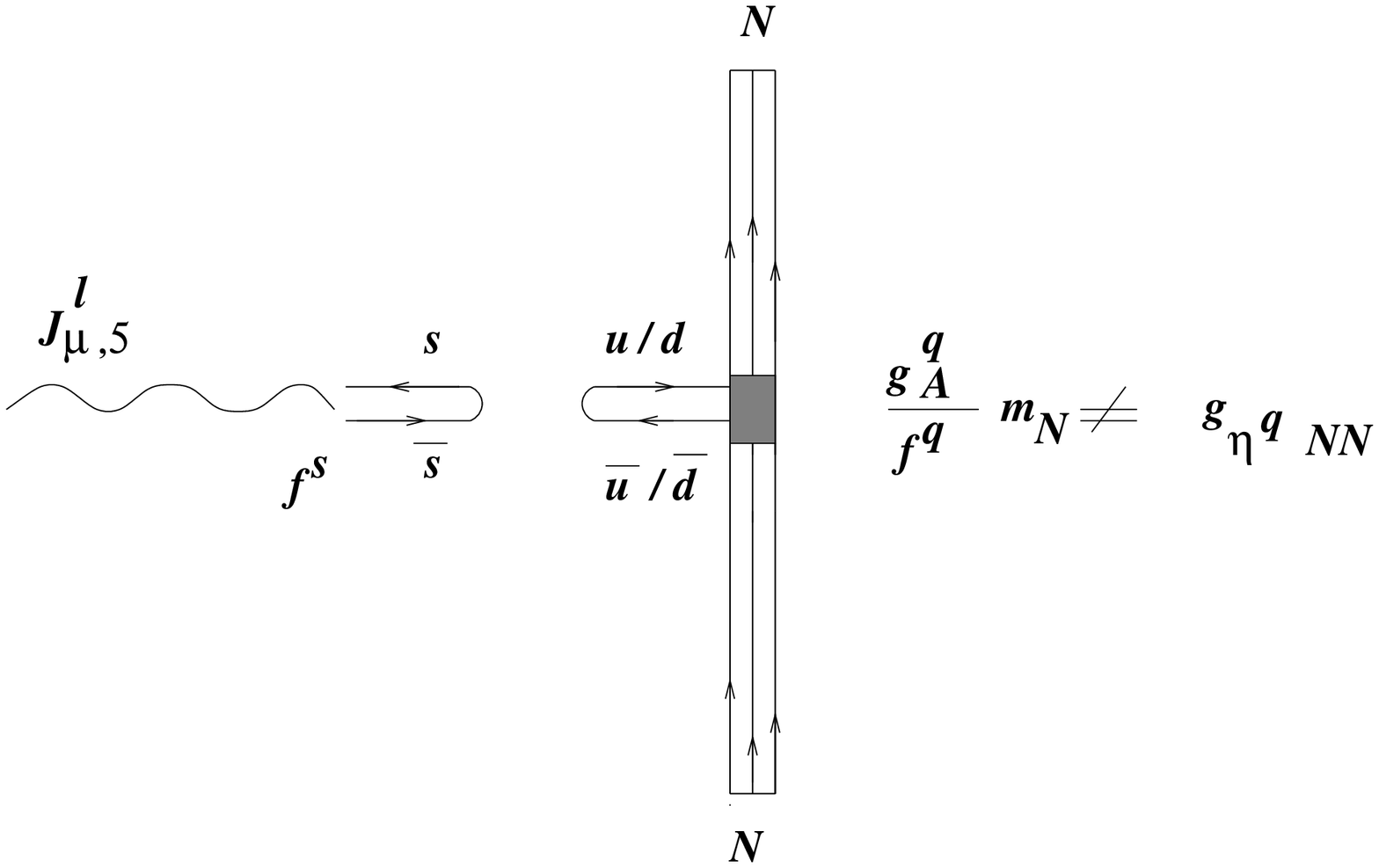,width=10cm}}
\vspace{0.1cm}
{\large Fig.\ 2\hspace{0.2cm} Violation of current universality by
the non-strange isosinglet axial current. }
\end{figure}

\begin{figure}[htbp]
\centerline{\psfig{figure=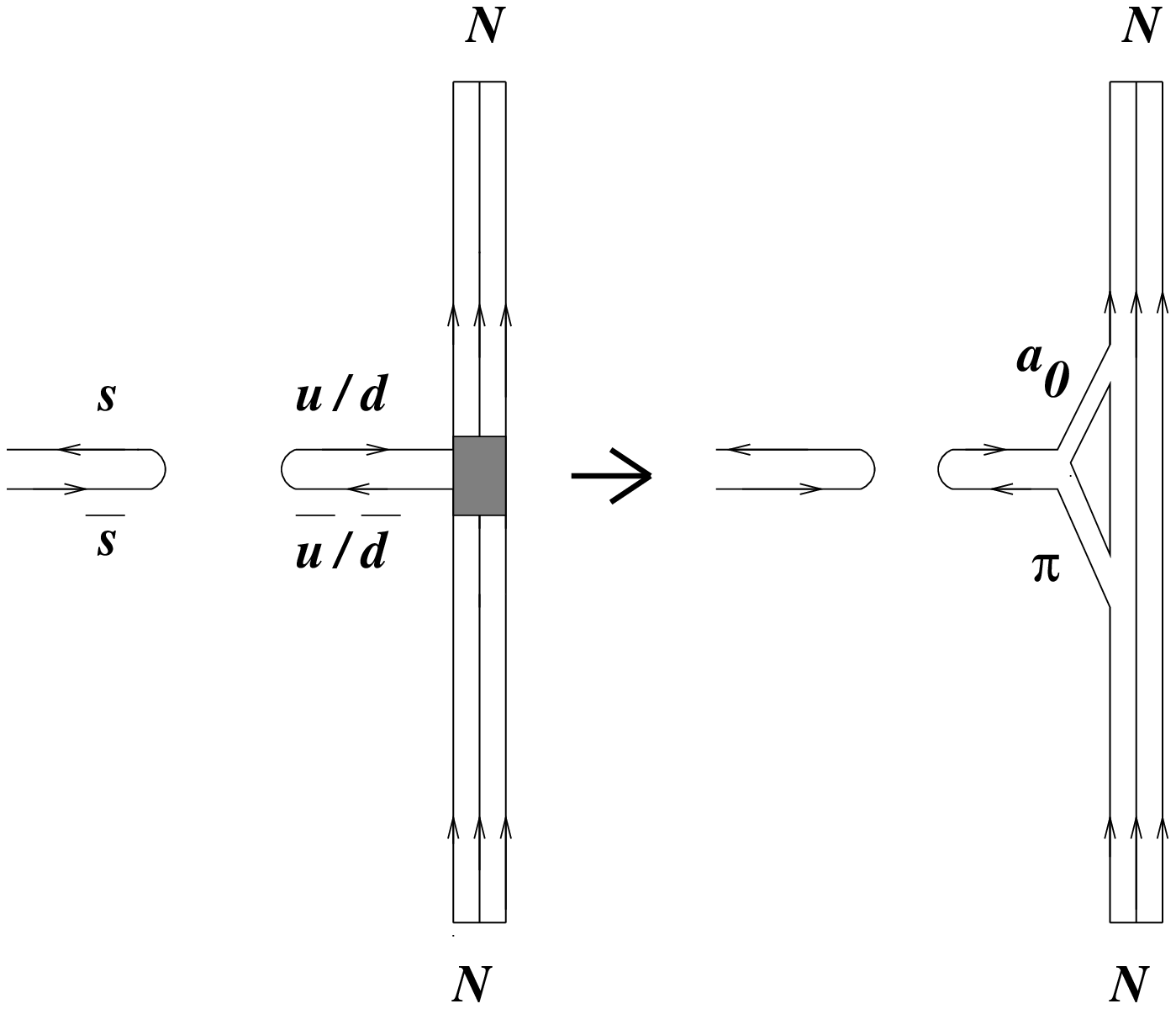,width=10cm}}
\vspace{0.1cm}
{\large Fig.\ 3\hspace{0.2cm} Coupling isosinglet $0^-$ non-strange
quarkonium to the nucleon via triangular loops. }
\end{figure}

\begin{figure}[htbp]
\centerline{\psfig{figure=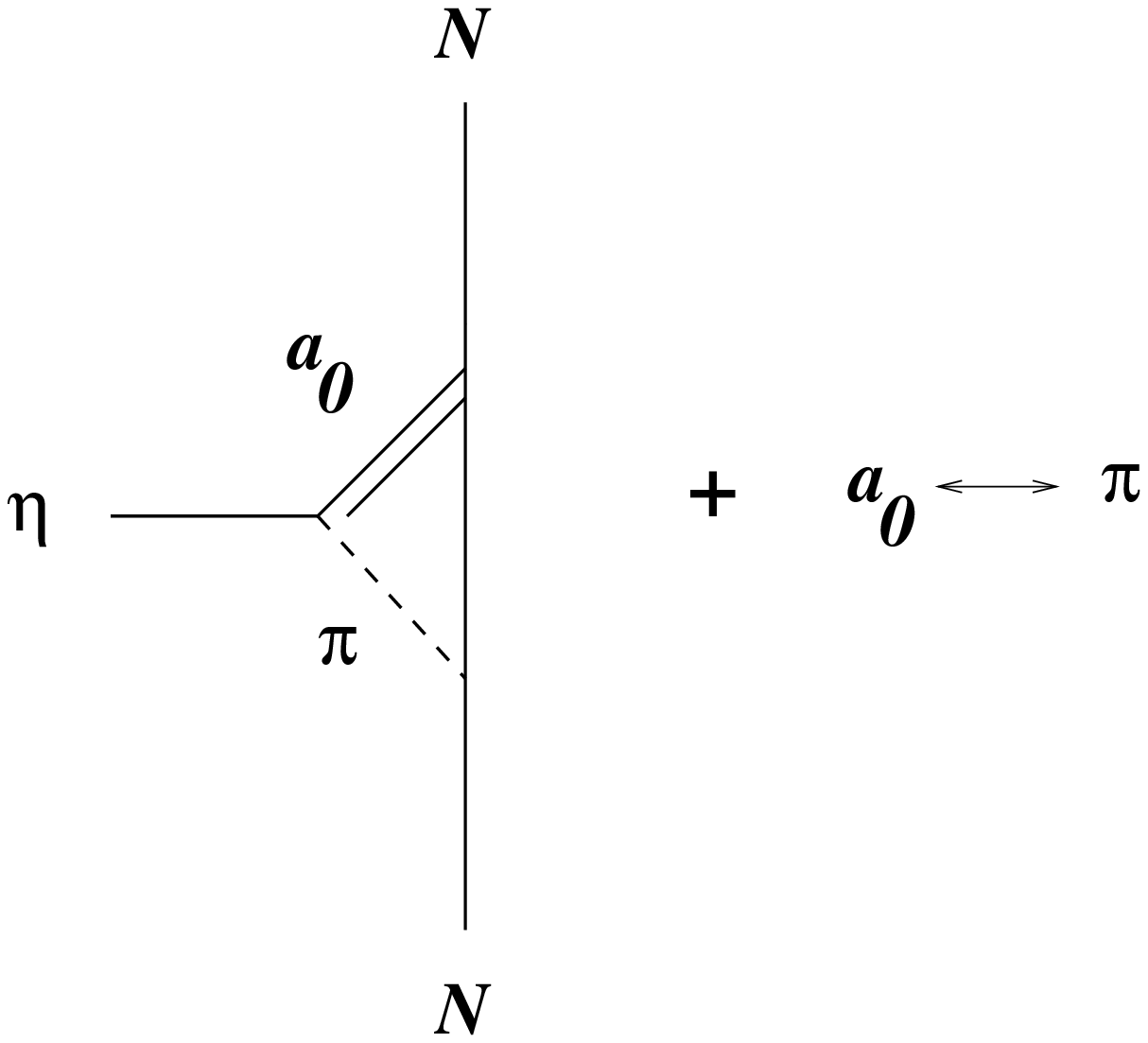,width=5cm}}
\vspace{0.1cm}
{\large Fig.\ 4\hspace{0.2cm} The triangular $\eta NN$ vertex. }
\end{figure}

\begin{figure}[htbp]
\centerline{\psfig{figure=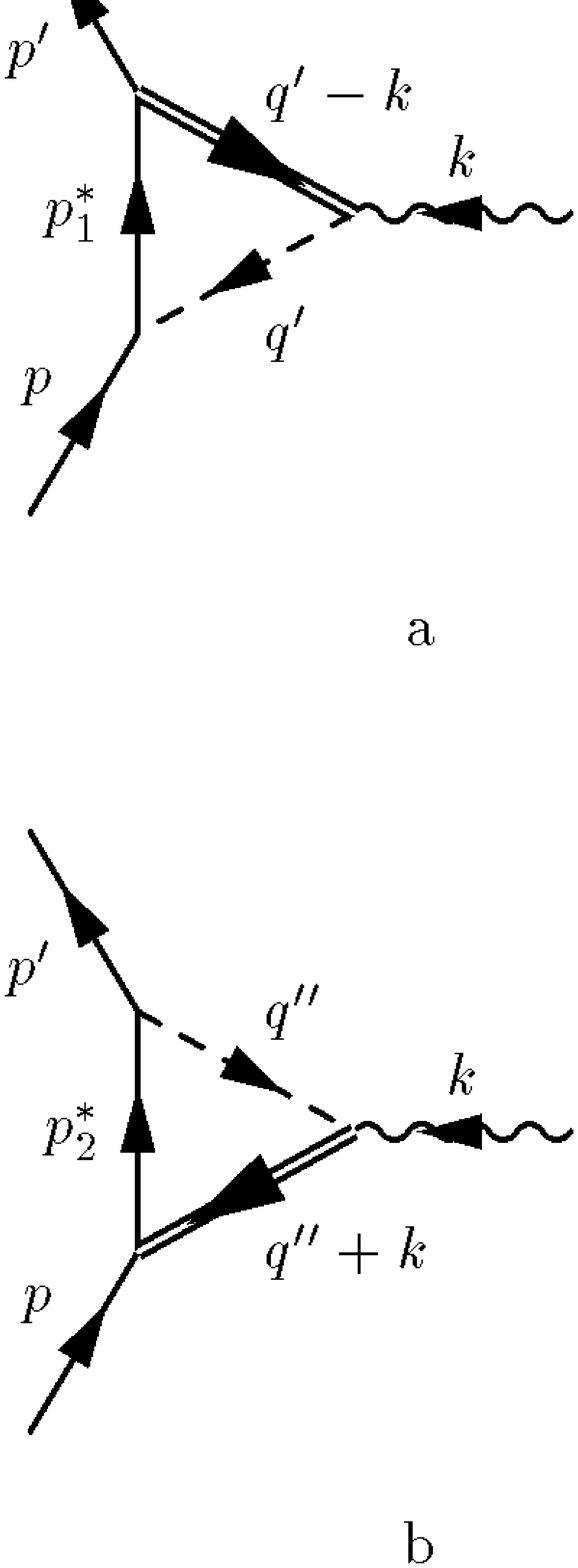,width=7cm}}
\vspace{0.1cm}
{\large Fig.\ 5\hspace{0.2cm} The flow of momentum.  }
\end{figure}

\begin{figure}[htbp]
\centerline{\psfig{figure=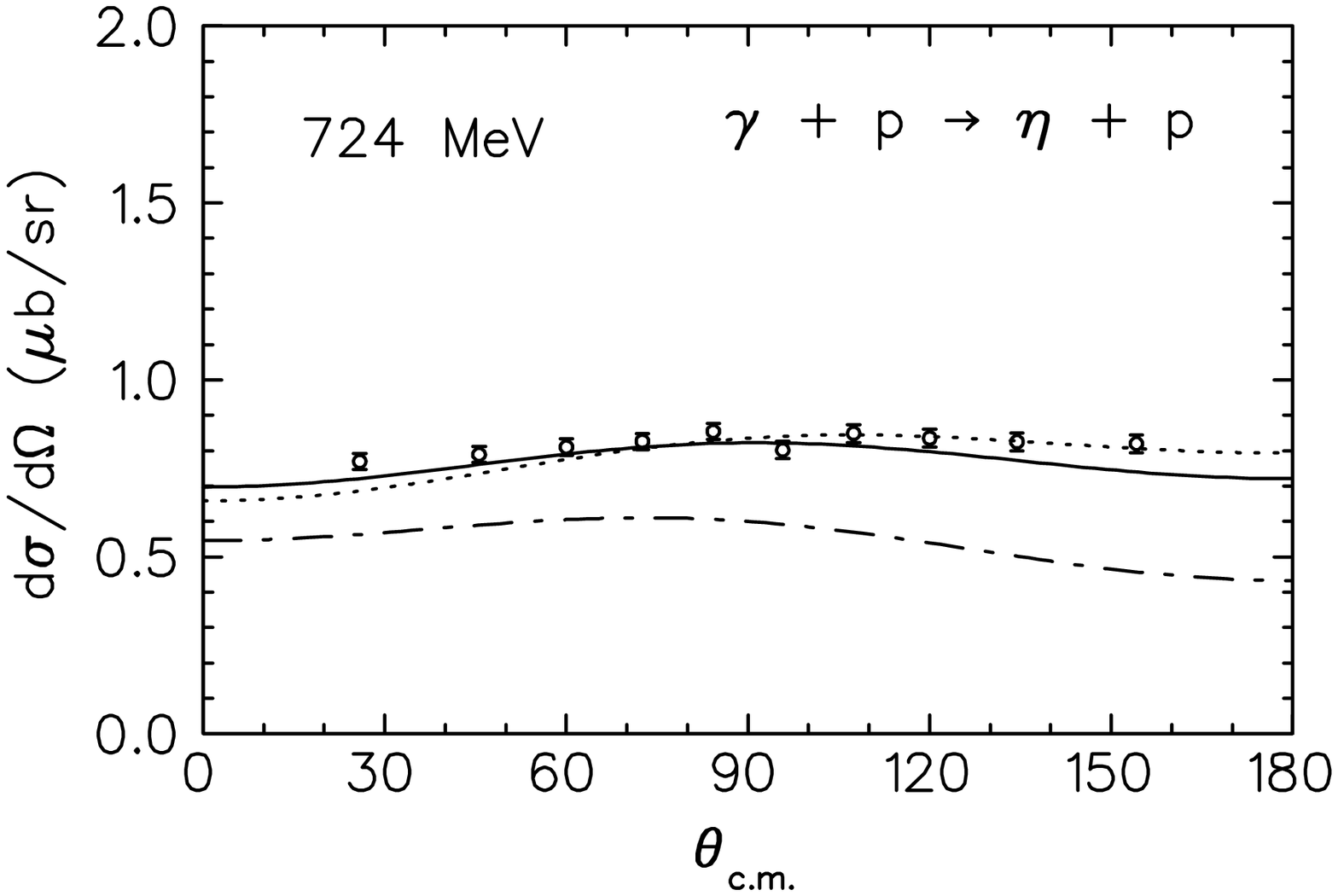,width=10cm}}
\vspace{0.1cm}
{\small Fig.\ 6\hspace{0.2cm}
Differential cross section for $\eta $ photo-production
off proton at lab energy of 724 MeV as calculated
within the model of Tiator, Bennhold and Kamalov ~\cite{TiKa}.
The dotted and dash--dotted lines correspond to 
$g_{\eta NN}^2/4\pi $ taking the values of 0.4 and 1.1, respectively.
The full line corresponds to the re-examined $a_0\pi N$ triangular 
coupling with $g_{a_0NN}^2/4\pi \approx $ 6.79. 
The data are taken from Krusche et al. \cite{TiKa}.
 }
\end{figure}

\begin{figure}[htbp]
\centerline{\psfig{figure=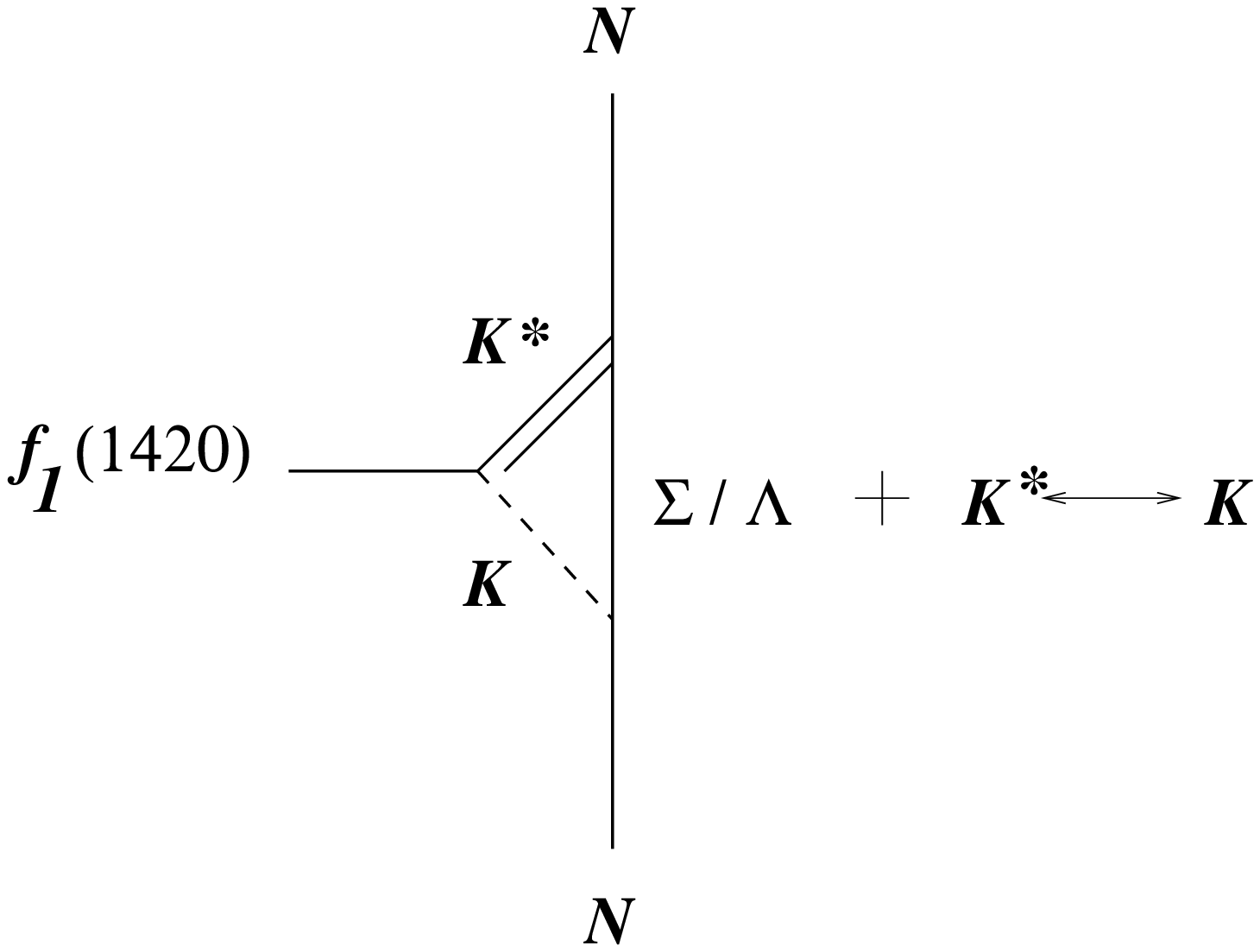,width=7cm}}
\vspace{0.1cm}
{\large Fig.\ 7\hspace{0.2cm}The triangular $f_1(1420) NN$ vertex.
}
\end{figure}


\newpage
\begin{thebibliography}{99}
\bibitem{KiWe}  M.\ Kirchbach and H.-J.\ Weber, Comm.\ Nucl.\ Part.\ Phys.\
                {\bf 22}, 171 (1998). 
\bibitem{Szczurek} H.\ Holtman, A.\ Szczurek, and J.\ Speth,
                   Nucl.\ Phys.\ {\bf A596}, 631 (1996).
\bibitem{Sav}  M.\ J.\ Savage and J.\ Walden, Phys.\ Rev.\ {\bf D55}, 
               5376 (1997). 
\bibitem{Feld} T.\ Feldmann, P.\ Kroll and B.\ Stech, Phys.\ Rev.\ {\bf D58},
               114006 (1998).
\bibitem{Ven} G.\ M.\ Shore and G,\ Veneziano, Phys.\ Lett.\ {\bf B244}, 
              75 (1990); Nucl.\ Phys.\ {\bf B381}, 23 (1992).
\bibitem{Feld1} T.\ Feldmann, Int.\ J.\ Mod.\ Phys.\ {\bf A}, 159 (2000).
\bibitem{proton} Spin Muon Collaboration (D.\ Adams {\it et al.\/} )
                 Phys.\ Rev.\ {\bf D56}, 5330 (1997).
\bibitem{Jaffe} R.\ L.\ Jaffe, Phys.\ Lett.\ {\bf B313}, 131 (1992).
\bibitem{PDG94} Review of Particle Properties, C.\ Caso {\it et al}., 
                 Eur.\  Phys.\ J.\  {\bf C3} (1998).
\bibitem{KiTi} M.\ Kirchbach and L.\ Tiator, Nucl.\ Phys.\ {\bf A604}, 
                385 (1996).
\bibitem{deAlf} V.\ De Alfaro, S.\ Fubini, G.\ Furlan, and C.\ Rossetti,
                {\it Currents in Hadron Physics}
                (Amsterdam: North Holland, 1973 ).\\
                 R.\ K.\ Bhaduri, {\it Models of the Nucleon\/}
                 (California, Addison--Wesley, 1988).
\bibitem{Coon} S.\ A.\ Coon and M.\ D.\ Scadron, Phys.\ Rev.\ {\bf C42},
                2256 (1990).
\bibitem{BMa}   B.\ Machet, Int.\ J.\ Mod.\ Phys.\ {\bf A14}, 4003 (1990).
\bibitem{TiKa} B.\ Krusche et al., Phys.\ Rev.\ Lett.\ {\bf 74}, 
               3736 (1995);\\ 
               L.\ Tiator, C. Bennhold, and S. Kamalov, Nucl.\ Phys.\ 
               {\bf A580}, 455 (1994).
\bibitem{Reuber} A.\ Reuber, K.\ Holinde, and J.\ Speth, Nucl.\ Phys.\ 
                {\bf A570}, 541 (1991). 
\bibitem{Ecker} G.\ Ecker, J.\ Gasser, A.\ Pich, and E. de Rafael,
                Nucl.\ Phys.\ {\bf B321}, 311 (1989).
\bibitem{Ma89} R.\ Machleidt, Adv.\ Nucl.\ Phys.\ {\bf 19}, 189 (1989).
\bibitem{Els87} R.\ Machleidt, K.\ Holinde, and C.\ Elster,
                Phys.\ Rep.\ {\bf 149}, 1  (1987).
\bibitem{GrKr}  W.\ Grein and P.\ Kroll, Nucl.\ Phys.\ {\bf A338}, 332 (1980).
\bibitem{Gross} F.\ Gross, J.\ W.\ Van Orden, and K.\ Holinde, 
                Phys.\ Rev.\ {\bf C41}, R1909 (1990).
\bibitem{Bolton} T.\ Bolton et al., Phys.\ Lett.\ {\bf B279} 495 (1992).
\bibitem{Zou} Bing-song ZOU, {\tt hep-ph/9611238} 
\bibitem{Kmol} A.\ Reuber, K.\ Holinde, H.\ C.\ Kim, and J.\ Speth,
               Nucl.\ Phys.\ {\bf A608}, 243 (1996).
\bibitem{OZ}  V.\ I.\ Ogievetsky and B.\ M.\ Zupnik, Nucl.\ Phys. 
                {\bf B24}, 612 (1970).
\bibitem{SN} S.\ Neumeier, Diploma Thesis, TH Darmstadt, Germany, 1996, 
                   unpublished.
\bibitem{Saghai} T.\ Mizutani, C.\ Fayard, G.-H. Lamot, and B.\ Saghai,
                 Phys.\ Rev.\ {\bf C58}, 1551 (1998).
\end{thebibliography}
\end{document}